\documentclass[12pt]{article}
\usepackage{graphicx}

\begin{document}

\centerline{\sc HIGH-SPEED PHOTOMETRY OF THE ECLIPSING}
\centerline{\sc CATACLYSMIC VARIABLE 1RXS J180834.7+101041}

\bigskip

\bigskip

\centerline{\it John Southworth$^1$ and Christopher M.\ Copperwheat$^2$}
\centerline{\it $^1$\,Astrophysics Group, Keele University, Staffordshire, ST5 5BG, UK}
\centerline{\it $^2$\,Department of Physics, University of Warwick, Coventry, CV4 7AL, UK}

\bigskip

\bigskip

\noindent\parbox{\textwidth}{\large We present high-speed photometry covering one eclipse of the cataclysmic variable system 1RXS J180834.7+101041. The data are modelled as arising from a close binary system containing a low-mass stars filling its Roche lobe plus a white dwarf with an accretion disc. We find an orbital inclination of $77.07 \pm 0.47$ degrees and a mass ratio of $0.168 \pm 0.016$. The relatively low inclination means that only the accretion disc and not the white dwarf are eclipsed by the low-mass star in this system.}

\bigskip

\section*{Introduction}

The ROSAT satellite detected a source of X-ray radiation which appears in the {\it Bright Source Catalogue}$^1$ under the name 1RXS J180834.7+101041 (hereafter J1808+1010). As part of a systematic search for the optical counterparts of ROSAT sources, Denisenko et al.\ (ref.\,2) obtained time-resolved photometry of the region of sky containing this source, finding strong photometric variability for the star USNO-B1 1001-0317189. Their photometric monitoring lasted for 6.2\,hr and exhibits four sharp dimmings on a period of $0.070037 \pm 0.000001$ days. The star also showed longer-time brightness variations within the magnitude range 16.2--17.5. Based on these characteristics, Denisenko et al.\ classified J1808+1010 as a cataclysmic variable of possibly AM\,Her (magnetic$^3$) type.

% J1808+1010 was also detected by the {\it 2 Micron All Sky Survey} (2006AJ....131.1163S), finding magnitudes $J = 14.93 \pm 0.04$ and $K = 14.33 \pm 0.06$.

Follow-up observations of J1808+1010 were obtained by Bikmaev \& Sakhibullin$^4$. Their 11 phase-resolved low-resolution spectra showed strong hydrogen Balmer line emission and weaker helium line emission, which is a common characteristic of cataclysmic variables (hereafter CVs). The Balmer emission lines are clearly double-peaked, which is an indicator of a high orbital inclination. Bikmaev \& Sakhibullin also presented an $r$-band light curve covering 2.1\,hr at a time resolution of 25\,s, which showed a single sharp eclipse event of depth 0.8\,mag and total duration 8.6\,min. From their observations these authors confirmed that J1808+1010 is a CV, but favoured the SU UMa (non-magnetic) subtype of these objects.

CVs are a class of interacting binary star system which display a huge diversity of physical phenomena. The majority of them are composed of a white dwarf and a low-mass and unevolved secondary star, plus an accretion disc through which material passes from the secondary star to the white dwarf. This accretion disc often totally dominates the light of the system, so the only readily observable physical property of most CVs is its orbital period$^{5,6}$. The importance of eclipsing CVs lies in the information which can be extracted from them: detailed modelling of their eclipses allows one to obtain the basic physical properties of the system, including the masses and radii of the stellar components$^{7,8}$. Such information is valuable in understanding the evolution of CVs, and of other classes of interacting binary system, so we decided to obtain high-speed photometry of J1808+1010 to assess its usefulness for detailed study.

\section*{Observations}

\begin{figure} \includegraphics[width=\textwidth,angle=0]{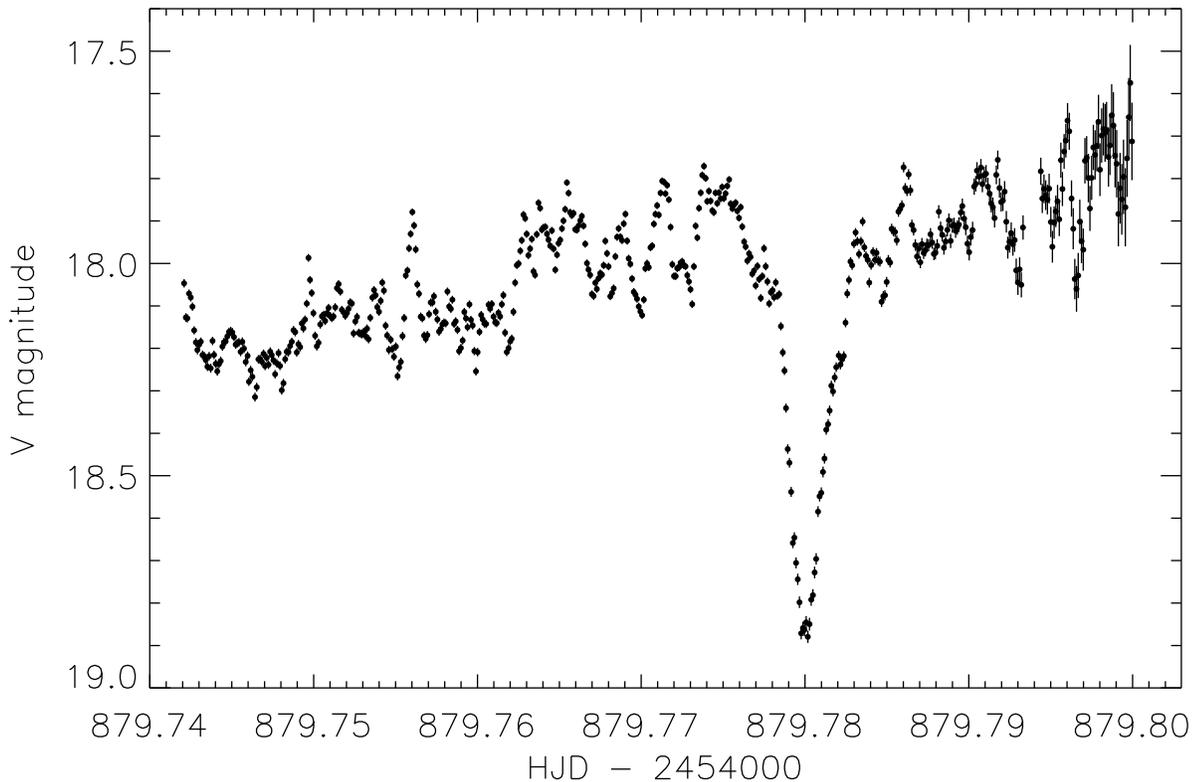}
\caption{\label{fig:plotLC} Plot of the full light curve obtained for
J1808+1010. In most cases the errorbars are smaller than the point size.}
\end{figure}

We observed J1808+1010 at the end of the night of 2009 February 16, starting at UT 05:56 and ending at 07:12. We used the Auxiliary Port focal station of the 4.2m William Herschel Telescope, operated by the Isaac Newton Group on the island of La Palma. A total of 555 observations were obtained through a $V$ filter. The detector was a 2148$\times$4200 pixel EEV CCD, windowed down to (2000\,px)$^2$ and binned 4$\times$4 to minimise the readout time. With an exposure time of 4\,s, we achieved a cadence of 9\,s. The data were reduced following standard procedures and using a pipeline written in {\sc idl}$^{9,10}$.

Our data are plotted in Fig.\,1, and cover one eclipse of depth 0.9\,mag and duration 6.5\,min. The variation in eclipse depth and duration compared to previous studies can be attributed to changes in the accretion disc size and structure$^{11}$. The outside-eclipse magnitude was $V \approx 18$, and the system was gradually increasing in brightness throughout our observing sequence. This puts it towards the faint end of the magnitude interval within which J1808+1010 has been seen, but this must be interpreted with caution due to the range of passbands used in previous observations. The substantial brightness variations outside eclipse are due to the phenomenon of flickering, which is caused by the stochastic nature of the mass transfer rate onto the accretion disc or white dwarf$^{12}$.

\section*{Analysis}

\begin{figure} \includegraphics[width=\textwidth,angle=0]{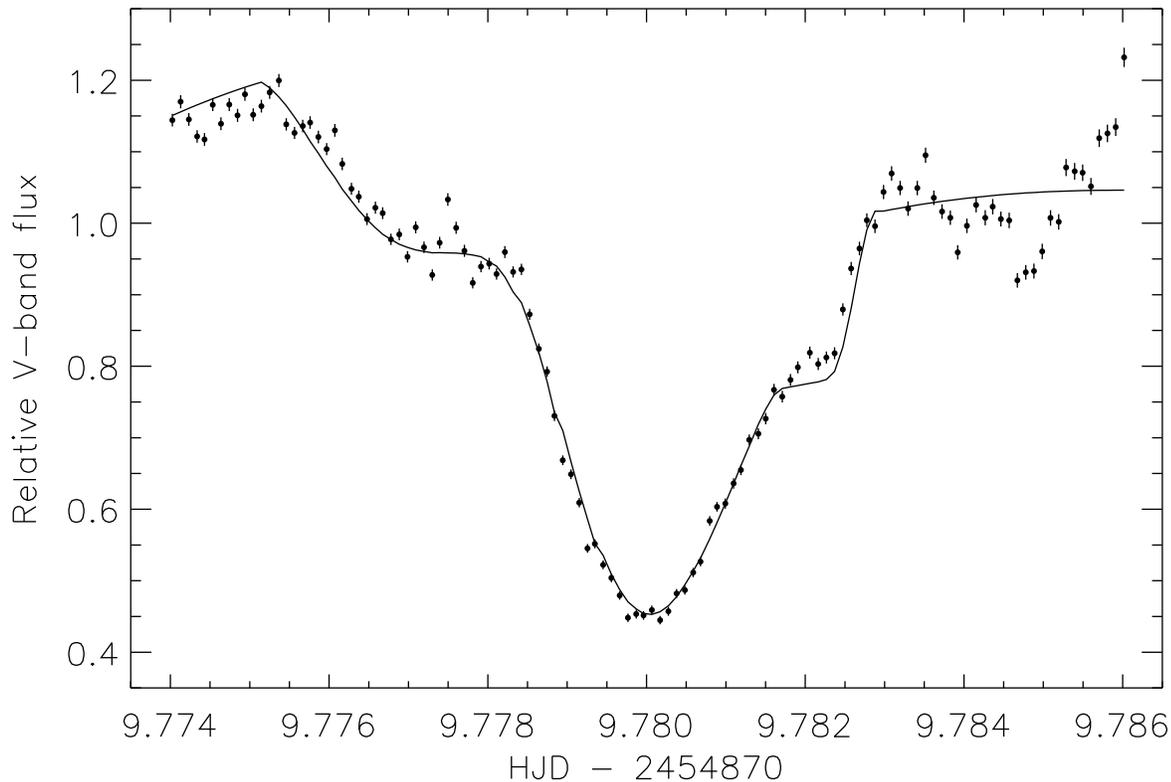}
\caption{\label{fig:plotLCfit} Comparison between the observed data for
J1808+1010 (points with errorbars) and the best fit found using the
{\sc lcurve} code (solid line).} \end{figure}

A model was fitted to the light curve using the {\sc lcurve} code written by T.\ R.\ Marsh. This code models a CV as four components: the white dwarf, the secondary star, an accretion disc surrounding the white dwarf, and bright spot where the mass transfer stream intersects with the edge of the accretion disc. A detailed description of {\sc lcurve} can be found in Copperwheat et al.\ (ref.\,13). For optimisation of the parameters of the model we used the Nelder-Mead downhill simplex$^{14}$, Levenberg-Marquardt$^{14}$ and Markov Chain Monte Carlo methods. {\sc lcurve} includes a large number of parameters which are needed to represent the eclipses of CVs observed with very high precision and time resolution$^{13}$. Our data are not of this quality so these parameters were fixed at values which we have found to be physically appropriate in detailed modelling of other eclipsing systems$^{\rm 13,15-18}$.

In our analysis we found that only part of the accretion disc of J1808+1010 is eclipsed; the white dwarf is not eclipsed at any point in the orbit. This means that the eclipse shape is rather uninformative, and allows us only to constrain the orbital inclination ($i = 77.07 \pm 0.47$ degrees), the ratio of the mass of the secondary star to that of the white dwarf ($q = 0.168 \pm 0.016$), and the midpoint of the eclipse (HJD[UTC] $2454879.27851 \pm 0.00006$). The best fit (Fig.\,2) accurately reproduces the data for most of the eclipse but does not fit the egress of the bright spot well. An improvement in this situation would require similar observations of a larger number of eclipses in order to allow the flickering to be averaged out.

If the white dwarf were eclipsed, the times and durations of the partial phases of the eclipse would allow the radius of the white dwarf to be determined, and therefore the mass and radius of both the white dwarf and secondary star by the imposition of a theoretical mass--radius relation for white dwarfs$^{7,8}$. In the case of J1808+1010 we are unable to obtain these constraints, which means it is not well suited to detailed eclipse modelling analyses.

\section*{Summary and conclusions}

1RXS J180834.7+101041 is a cataclysmic variable which is known to be eclipsing with a period of 100.85 minutes. We obtained high-cadence photometry of one eclipse event, finding that the accretion disc is eclipsed but that the white dwarf is not. We have fitted a parametric model to the light curve, and measured the orbital inclination ($i = 77.07 \pm 0.47$ degrees) and mass ratio ($q = 0.168 \pm 0.016$) of the system. Because the white dwarf is not eclipsed we are unable to obtain additional constraints from our data; we find that J1808+1010 is not a promising system for the study of eclipses in CVs.

Whilst the current paper was in preparation we became aware of an independent study of J1808+1010 by Yakin et al.\ (ref.\,19) which presents photometry and spectroscopy of this system. Our light curve is of substantially better time resolution, and the orbital inclination and mass ratio we find are in agreement with (but more precise than) the values they obtain: $i = 78 \pm 1.5$ degrees and $q = 0.18 \pm 0.05$. Yakin et al.\ also find that J1808+1010 has an asymmetric brightness distribution in its accretion disc (see ref.\,20 for an extreme example of this phenomenon). From their observations and from some simple relations Yakin et al.\ infer the masses of the two stars to be $0.8 \pm 0.22$ and $0.14 \pm 0.02$ M$_\odot$ for the white dwarf and secondary component, respectively.

\section*{References}

\begin{small}
$^1$ W.\ Voges, {\it et al.}, 1999, {\it A\&A}, {\bf 349}, 389. \\
$^2$ D.\ V.\ Denisenko, T.\ V.\ Kryachko, B.\ L.\ Satovskiy, 2008, {\it Astronomers Telegram} {\bf 1640}. \\
$^3$ B.\ Warner, 1995, {\it Cataclysmic Variable Stars} (Cambridge University Press). \\
$^4$ I.\ F.\ Bikmaev, N.\ A.\ Sakhibullin, 2008, {\it Astronomers Telegram} {\bf 1648}. \\
$^5$ J.\ Southworth, {\it et al.}, 2006, {\it MNRAS}, {\bf 373}, 687. \\
$^6$ J.\ Southworth, {\it et al.}, 2007, {\it MNRAS}, {\bf 382}, 1145. \\
$^7$ S.\ P.\ Littlefair, {\it et al.}, 2006, {\it Science}, {\bf 314}, 1578. \\
$^8$ S.\ P.\ Littlefair, {\it et al.}, 2008, {\it MNRAS}, {\bf 388}, 1582. \\
$^9$ J.\ Southworth, {\it et al.}, 2009, 2009, {\it MNRAS}, {\bf 396}, 1023. \\
$^{10}$ J.\ Southworth, {\it et al.}, 2009, 2009, {\it MNRAS}, {\bf 399}, 287. \\
$^{11}$ J.\ Shears, {\it et al.}, 2010, {\it BAAS}, in press (arXiv:1005.3219). \\
$^{12}$ A.\ Bruch, 2000, {\it A\&A}, {\bf 359}, 998. \\
$^{13}$ C.\ M.\ Copperwheat, {\it et al.}, 2010, {\it MNRAS}, {\bf 402}, 1842. \\
$^{14}$ W.\ H.\ Press, S.\ A.\ Teukolsky, W.\ T.\ Vetterling, B.\ P.\ Flannery, 1992, {\it Numerical recipes in FORTRAN 77. The art of scientific computing (2nd Edn.)} (Cambridge University Press). \\
$^{15}$ J.\ Southworth, {\it et al.}, 2009, {\it A\&A}, {\bf 507}, 929. \\
$^{16}$ J.\ Southworth, C.\ M.\ Copperwheat, B.\ T.\ G\"ansicke, S.\ Pyrzas,, 2010 {\it A\&A}, {\bf 510}, A100. \\
$^{17}$ S.\ Pyrzas, {\it et al.}, 2009, {\it MNRAS}, {\bf 394}, 978. \\
$^{18}$ A.\ Nebot G\'omez-Mor\'an, {\it et al.}, 2009, {\it A\&A}, {\bf 495}, 561. \\
$^{19}$ D.\ G. Yakin, {\it et al.}, 2010, in K.\ Werner \& T.\ Rauch (eds.), {\it 27th European White Dwarf Workshop} (AIP Conference Proceedings), {\bf 1273}, 346. \\
$^{20}$ J.\ Southworth, {\it et al.}, 2010., {\it A\&A}, {\bf 524}, A86 \\
\end{small}

\end{document}